\documentclass[sigconf,natbib=true]{acmart}
\AtBeginDocument{%
  }

\usepackage{algorithm}
\usepackage{algorithmic}
\usepackage{amsfonts}
\usepackage{amsmath}
\usepackage{booktabs}
\usepackage{array}
\usepackage{colortbl}
\newcolumntype{R}[1]{>{\raggedleft\arraybackslash}p{#1}}  
\usepackage{multirow}

\setcopyright{acmlicensed}
\copyrightyear{2018}
\acmYear{2018}
\acmDOI{XXXXXXX.XXXXXXX}
\acmConference[Conference acronym 'XX]{Make sure to enter the correct
  conference title from your rights confirmation email}{June 03--05,
  2018}{Woodstock, NY}
\acmISBN{978-1-4503-XXXX-X/2018/06}

\begin{document}

\title{RecThinker: An Agentic Framework for Tool-Augmented Reasoning in Recommendation}






\author{Haobo Zhang}
\author{Yutao Zhu}
\affiliation{%
  \institution{Gaoling School of Artificial Intelligence}
  \institution{Renmin University of China}
  \city{Beijing}
  \country{China}
}
\email{{zhanghb,ytz}@ruc.edu.cn}

\author{Kelong Mao}
\author{Tianhao Li}
\affiliation{%
  \institution{JD.com}
  \city{Beijing}
  \country{China}
}
\email{maokelong.1@jd.com}

\author{Zhicheng Dou}
\authornote{Corresponding author}
\affiliation{%
  \institution{Gaoling School of Artificial Intelligence}
  \institution{Renmin University of China}
  \city{Beijing}
  \country{China}}
\email{dou@ruc.edu.cn}

\renewcommand{\shortauthors}{Trovato et al.}

\begin{abstract}
Large Language Models (LLMs) have revolutionized recommendation agents by providing superior reasoning and flexible decision-making capabilities.
However, existing methods mainly follow a passive information acquisition paradigm, where agents either rely on static pre-defined workflows or perform reasoning with constrained information. It limits the agent's ability to identify information sufficiency, often leading to suboptimal recommendations when faced with fragmented user profiles or sparse item metadata.
To address these limitations, we propose RecThinker, an agentic framework for tool-augmented reasoning in recommendation, which shifts recommendation from passive processing to autonomous investigation by dynamically planning reasoning paths and proactively acquiring essential information via autonomous tool-use. 
Specifically, RecThinker adopts an \textit{Analyze-Plan-Act} paradigm, which first analyzes the sufficiency of user-item information and autonomously invokes tool-calling sequences to bridge information gaps between available knowledge and reasoning requirements.
We develop a suite of specialized tools for RecThinker, enabling the model to acquire user-side, item-side, and collaborative information for better reasoning and user-item matching. 
Furthermore, we introduce a self-augmented training pipeline, comprising a Supervised Fine-Tuning (SFT) stage to internalize high-quality reasoning trajectories and a Reinforcement Learning (RL) stage to optimize for decision accuracy and tool-use efficiency.
Extensive experiments on multiple benchmark datasets demonstrate that RecThinker consistently outperforms strong baselines in the recommendation scenario.

\end{abstract}



\received{20 February 2007}
\received[revised]{12 March 2009}
\received[accepted]{5 June 2009}

\maketitle

\section{Introduction}
\begin{figure}[t]
\centering
\includegraphics[width=1.0\linewidth]{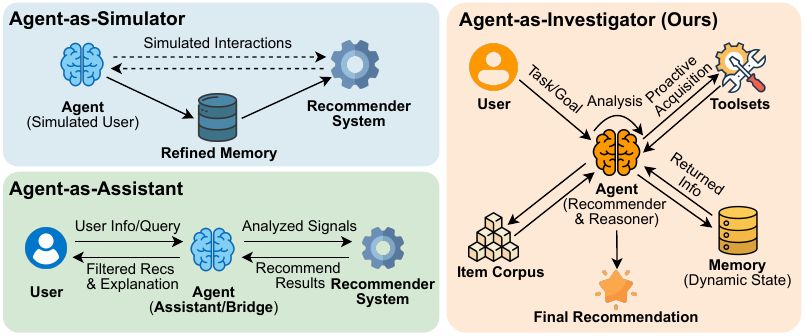}
\caption{Illustration of agentic recommendation paradigms.
}
\label{intro_graph}
\end{figure}

Recommender systems (RS), which aim to filter and rank items that best match users' preferences from large candidate pools, have become fundamental components of modern information systems. 
With the advances of LLMs, researchers have increasingly explored their application in RS. 
Capitalizing on LLMs' rich world knowledge and strong semantic understanding, studies aim to address complex user needs.
Early work~\cite{InstructionLLMRec2023, LLMRank2024, TALLRec, LlamaRec} attempted to employ LLMs directly as recommenders through specialized prompting and fine-tuning on user–item interaction data. 
Subsequently, researchers~\cite{KAR2023, LLMRec2024, UR4Rec2025} have explored using LLMs to enhance traditional RS by generating richer user and item representations. 
However, these paradigms often struggle with issues such as hallucinations, semantic misalignment between the knowledge space of LLMs and the RS's objective, and incomplete utilization of the LLM's inherent reasoning capabilities.
More recently, the emergence of agentic recommendation~\cite{AgentCF2023, Agent4rec2024} has introduced a new paradigm. Driven by advancements in reasoning models, this approach equips LLM-powered agents with reasoning and interaction capabilities to analyze user information and guide the recommendation process.
Unlike conventional models, agentic recommenders mainly use a reasoning agent that analyzes user information, interacts with environments, and refines its internal model for better recommendations. 
Existing research in this field can be categorized into three paradigms. First, Agent-as-Simulator~\cite{AgentCF2023, RecAgent2025, Agent4rec2024}, focuses on simulating user-item interactions and their interactions with environments. In this paradigm, agents are constructed to simulate real-world user behaviors, to refine user profiles and memory systems to improve recommendation models. 
The second paradigm is Agent-as-Assistant~\cite{RAH2024, BiLLP2024, MACRS2024}, where the agent serves as an intelligent bridge between the user and the RS. Agents utilize their reasoning capabilities to analyze information provided by users, produce signals for recommendations, and iteratively adjust their strategies based on user feedback. For example, RAH~\cite{RAH2024} leverages agents to analyze user inputs, perform actions on items, and refine a personality library through a reflection mechanism.  Further studies~\cite{PersonaX2025, ToolRec2024, KGLA2024} improve user interest modeling and analytical performance through designing specialized components, such as fine-grained profilers, memory modules, or tools for candidate filtering. However, these systems mostly operate within a predefined information scope and perform passive comprehension and integration, resulting in static and limited information acquisition. 
The third paradigm, Agent-as-Investigator~\cite{RecMind2024}, deploys the agent directly as the core recommender, performing end-to-end reasoning, information acquisition, analysis, and recommendation.
Instead of passively receiving data, this approach proactively explores and analyzes tasks to complete recommendations. RecMind~\cite{RecMind2024} represents an early exploration of this paradigm, demonstrating the potential of agents to independently guide the recommendation.
Despite its potential, the investigator paradigm faces significant challenges.
First, there is a lack of explicit assessment regarding information sufficiency. While agents can interact with external tools, their information acquisition behaviors remain largely uncertain and opportunistic. 
Such behaviors are primarily driven by intermediate states rather than information deficiency analysis. These methods rarely analyze the information gap between available data and the requirements for accurate recommendations, leading to ineffective tool use. 
Consequently, these agents struggle to accurately perform user-item matching, resulting in suboptimal reasoning.
Second, the tool design in current methods remains limited. Many frameworks focus solely on retrieval and ranking tools to narrow the candidate pool, which offers little external knowledge to support deeper reasoning. Some works rely on generic search tools that are not specifically tailored for recommendation scenarios. As a result, the information acquired is often incomplete and one-sided, failing to construct a deep reasoning chain and making comprehensive decisions based on multi-dimensional evidence.
Moreover, existing investigator agents lack mechanisms for policy evolution. Most current frameworks rely on static policies or fixed prompts to guide tool invocation and analysis. They do not evolve their strategies based on the complexity of the task or the specific user environment. It limits their ability to explore the reasoning spaces, making it difficult to handle complex scenarios where the information gap is significant and requires sophisticated investigation.

To address these challenges, we argue that effective recommendation agents should analyze the gap between available information and the evidence required for accurate decisions. Furthermore, they should proactively acquire multi-dimensional evidence through specialized tools and continuously optimize their policy to handle diverse recommendation scenarios.
Therefore, we propose RecThinker, an agentic framework for tool-augmented reasoning in recommendations.
As an investigator agent, it is capable of autonomously analyzing recommendation tasks, identifying the information required for reasoning, and actively obtaining missing evidence through flexible tool invocation.
Specifically, RecThinker adopts an \textit{Analyze-Plan-Act} workflow. Instead of passively processing inputs, the agent first assesses the sufficiency of available user and item information. It then strategically plans and executes tool-calling sequences to acquire missing evidence and bridge the information gap. 
To support this process, we develop a set of specialized tools for recommendation scenarios to support user preference analysis, item knowledge completion, and collaborative information acquisition. These tools empower the agent to synthesize diverse evidence and perform more accurate user–item matching. 
Furthermore, to improve the precision of reasoning and reduce redundant tool-use, we introduce a two-stage training strategy to continuously optimize the policy.  We first employ a self-augmentation approach to fine-tune the model using high-quality reasoning trajectories. 
Subsequently, we apply reinforcement learning to optimize the agent's policy for a deeper and more accurate investigation and more effective tool-usage.
Through these components, RecThinker empowers agents to reason more effectively in complex environments and deliver more accurate and transparent recommendations.
We conduct comprehensive experiments on multiple real-world datasets, demonstrating the superiority of RecThinker over strong baselines across diverse recommendation scenarios.

In summary, our main contributions are as follows:

(1) We propose RecThinker, an agentic framework for reasoning with tool augmentation in recommendation, which analyzes recommendation tasks autonomously and acquires necessary evidence proactively via flexible tool invocation.

(2) We introduce an Analyze-Plan-Act reasoning paradigm that enables the agent to assess information sufficiency, plan appropriate tool usage, and iteratively refine its reasoning process for recommendation.

(3) We develop a suite of specialized tools for user preference analysis, item information completion, and collaborative information acquisition in recommendation scenarios.

(4) We introduce a two-stage training strategy that combines self-augmented SFT on high-quality trajectories with RL to improve reasoning accuracy and tool invocation effectiveness.

\section{Related Work}

\subsection{Agent-based Recommendation}

Agent-based recommendation leverages the reasoning and planning abilities of LLMs to model recommendations as autonomous decision-making tasks.
Recent studies can be broadly categorized into three paradigms based on the agent's role and interaction logic.

\textbf{Agent-as-Simulator}. This paradigm employs agents to simulate interactions between users, items, and the environment. These methods~\cite{RecAgent2025, SimUSER2025, AgentCFplus2025, CSHI2025} mainly replicate real-world behaviors such as clicks and purchases, and gradually refine agent memories. 
For example, AgentCF~\cite{AgentCF2023} models both users and items as agents and simulates their interactions to train the agents and capture collaborative signals.
Agent4Rec~\cite{Agent4rec2024} equips agents with profile, memory, and action modules to simulate complex behaviors such as browsing and page-flipping.
These works mainly focus on comprehensive user behavior modeling and interaction simulation, while paying limited attention to integrating the agents with RSs.

\textbf{Agent-as-Assistant}. In this paradigm, agents act as an intelligent assistant between the user and RS. Benefiting from the advancement of reasoning techniques, these agents leverage LLM reasoning to process user-provided information, infer recommendation signals, and refine their behaviors through feedbacks.~\cite{RAH2024, BiLLP2024, MACRS2024}
For example, RAH~\cite{RAH2024} conducts a fine-grained analysis of user personalities, performs actions on items, and refines actions and personality library based on the feedback. 
Subsequent studies~\cite{RuleAgent2025, LLM4Rerank2025, InteRecAgent2025, KGLA2024} focused on designing more specialized components to improve the agent’s analytical performance and user preference capture.
For example, PersonaX~\cite{PersonaX2025} designs a specialized profiler that segments and clusters long behavior sequences and selects sub-behaviors to support real-time profile generation. InteRecAgent~\cite{InteRecAgent2025} designs a long- and short-term memory and a candidate memory bus to model structured profiles and narrow candidate pools by using recommenders as tools.
Despite these advances, these agents mainly operate within a predefined information scope. Their reasoning processes are largely constrained to passive understanding and static and limited information acquisition.

\textbf{Agent-as-Investigator}. In this paradigm, the agent serves as the core recommender that actively conducts multi-step reasoning and exploration. Unlike previous paradigms, these methods enable agents to autonomously analyze tasks, proactively interact with tools, and explore the environment to find evidence. 
RecMind~\cite{RecMind2024} represents a pioneering effort that evolves CoT into a self-inspiring mechanism to plan and guide tasks, and utilizes search tools to assist the agent.
However, existing methods still struggle with systematic information gap analysis and often rely on generic toolsets. 
In this work, we propose RecThinker, which advances this paradigm by considering the information insufficiency of users and items and developing a recommendation-specific toolset for better reasoning and ranking.

\subsection{LLM Reasoning for Recommendation}
With the wide application of LLMs in RS, reasoning capability has become increasingly important for accurate recommendations. Early studies~\cite{LLMRank2024, COTRec2025, HetGCoT-Rec} attempted to incorporate chain-of-thought (CoT) techniques~\cite{COT2022} into recommenders, enabling LLMs to think step-by-step before producing final results. For example, InstructRec~\cite{InstructionLLMRec2023} incorporates CoT-based reasoning and instruction tuning to enhance recommendations. However, these prompt-based CoT methods cannot fully exploit the reasoning ability of LLMs.
The recent emergence of test-time computation and Large Reasoning Models (LRM), such as OpenAI-o1~\cite{openaio12024} and DeepSeek-R1~\cite{DeepSeekR1}, has significantly enhanced the LLM reasoning.
Consequently, subsequent research has begun to incorporate reasoning into RS. 
For instance, Reason4Rec~\cite{Reason4Rec2025} designs multiple experts to perform preference distillation, matching, and rating prediction through multi-step reasoning. 
Furthermore, several works~\cite{Reason2Recommend2025, DeepRec2025} utilize RL techniques, such as GRPO~\cite{GRPO2024}, to further enhance the reasoning ability. 
For example, R2Rec~\cite{R2ec2025} extends GRPO and designs RecPO to jointly optimize reasoning and recommendations.
Moreover, reasoning has become the core of agents, where LLMs are used for task analysis, planning, and decision control through explicit reasoning processes. In this work, we leverage LRM as the agent to autonomously analyze tasks and proactively invoke appropriate tools based on information needs, thereby improving recommendation performance.

\begin{figure*}[t]
\centering
\includegraphics[width=0.95\linewidth]{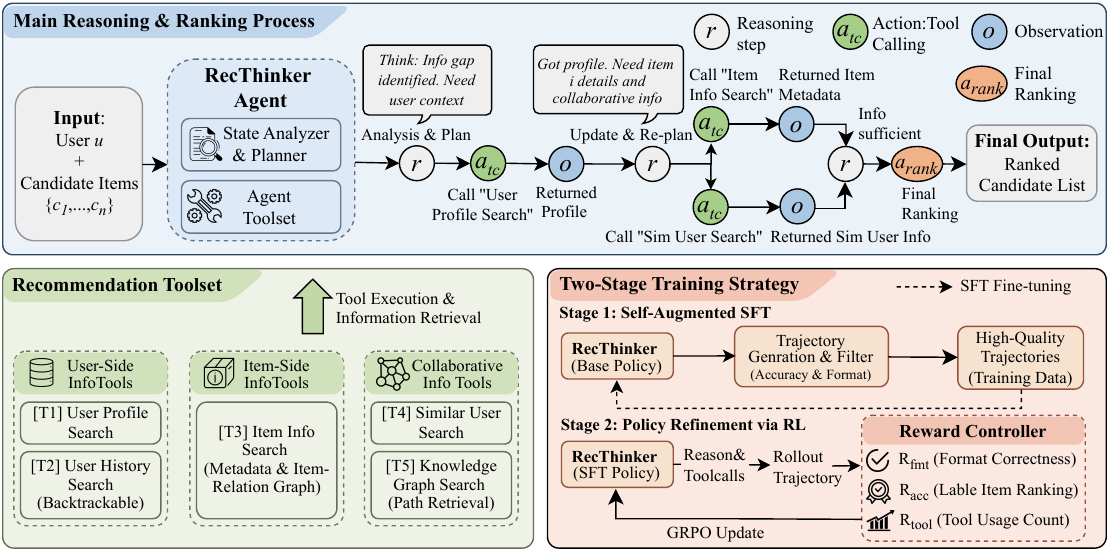}
\caption{The overall architecture of our RecThinker model. }
\label{model_graph}
\end{figure*}

\section{Method}
In this section, we propose RecThinker, a tool-augmented agentic recommendation framework that autonomously analyzes recommendation tasks and information gaps, and proactively acquires necessary information through flexible tool invocation. The overall architecture of RecThinker is shown in Figure~\ref{model_graph}.
We first formulate the agentic recommendation task, and then introduce the reasoning workflow and core architecture of RecThinker. Next, we describe the design of specialized tools and the information acquisition strategy. Finally, we present the proposed two-stage training paradigm, including self-augmented supervised fine-tuning (SFT) and reinforcement learning (RL) optimization.

\subsection{Problem Formulation}
In this paper, we denote the sets of users and items in the RS as $U$ and $I$, and their sizes as $N_U$ and $N_I$, respectively.
The user's behavior sequence will be organized in chronological order and represented as $S_u = \{i_{1},..., i_{n}\}, \forall u \in U$. 
In this work, we take the LLM as the core component and central controller for recommendation. 
We focus on a ranking task, where an LLM-based agent is employed to rank a set of candidate items $\{c_{1},..., c_{n}\}$ for a user $u$: $f_{\text{agent}}(u, \{c_{1},..., c_{n}\})$.
Given a user $u$ and a candidate item set $\{c_1,\ldots,c_n\}$, the agent iteratively performs reasoning and tool invocation to acquire auxiliary information and refine its understanding of user preferences.
Formally, the agent conducts a sequence of reasoning steps $\{r_1,\ldots,r_T\}$ and corresponding tool calls $\{a_1,\ldots,a_T\}$, where each $a_t$ is selected from a predefined tool set $\mathcal{A}$.
Based on the accumulated knowledge, the agent summarizes user preferences and matches them with candidate items to produce the final ranking:
\begin{equation}
    c_{1}',..., c_{n}' = f_{\text{agent}}(u, \mathcal{A}, \{c_{1},..., c_{n}\}),
\end{equation}
where $f_{\text{agent}}$ represents the overall decision process realized by iteratively executing a policy $\pi$ over multiple reasoning turns.

\subsection{Analyze-Plan-Act Workflow}

We propose an Analyze-Plan-Act workflow to structure the agent's decision-making process.
To provide a systematic formulation for RecThinker, we formalize this workflow as a multi-step reasoning trajectory. Given a user $u$ and a candidate set $\mathcal{C}=\{c_1,\dots,c_n\}$, the agent generates a sequence of states, actions, and observations. The recommendation process is modeled as a sequence of $T$ reasoning turns. At each turn $t \in \{1, \dots, T\}$, the agent's state is represented by the cumulative trajectory $\tau_t = (r_1, a_1, o_1, \dots, r_{t-1}, a_{t-1}, o_{t-1})$, where $r$ denotes internal reasoning, $a$ denotes the action (tool invocation), and $o$ represents the external observation returned by tool executions, which provides knowledge related to the user and items to support subsequent reasoning and ranking.
The agent's behavior is determined by a policy $\pi$, which generates the next thought and action based on the current trajectory:
\begin{equation}
    (r_t, a_t) \sim \pi(r_t, a_t \mid u, \mathcal{C}, \tau_t; \Theta),
\end{equation}
where $\Theta$ is the parameters of the agent, $r_t$ denotes an internal reasoning step that analyzes the current information context and identifies missing evidence, and $a_t \in \mathcal{A} \cup \{\text{RANK}\}$ is the action selected from the external tool set $\mathcal{A}$ or the final ranking operation.


The core of the policy $\pi$ lies in its ability to conduct information sufficiency assessment during multi-step reasoning. At each turn $t$, the agent systematically analyzes two categories of information: (1) User-Centric Information ($K_u$): The information of the user's preferences, short-term and long-term profile, and inferred user intent. (2) Item-Centric Knowledge ($K_{c_i}$): The attribute descriptions, semantic properties, and contextual relations of each candidate item $c_i \in \mathcal{C}$. 
The reasoning step $r_t$ performs information gap analysis by identifying missing or insufficient evidence required for reliable decision making. Formally, the information gap $\Delta_t$ is defined as:
\begin{equation}
    \Delta_t = f_{\mathrm{a}}(K_u, \{K_{c_i}\}_{i=1}^n),
\end{equation}
where $f_{\mathrm{a}}$ represents an assessment process implemented by the agent to compare $K_u$ with $\{K_{c_i}\}$ and assess whether the available information is sufficient for accurate ranking.

Based on the identified gap $\Delta_t$, the agent follows a decision-making procedure consisting of three stages: (1) Information Sufficiency Check. If $\Delta_t \approx \emptyset$, indicating that the available evidence is sufficient to support reliable ranking over all candidates, the agent takes the action of final ranking $a_t = \text{RANK}$.
(2) Proactive Information Acquisition. If critical information is missing, such as unclear user preferences toward specific attributes, or insufficient item descriptions, the agent formulates a plan to invoke one or multiple tools $a_t \in \mathcal{A}$ to acquire complementary evidence. 
(3) Observation Feedback: Each action $a_t$ returns an observation $o_t$, which includes extra information of users and items and provides additional information gain. The agent then updates the trajectory $\tau_{t+1}$ by integrating $o_t$ into its reasoning state, effectively refining its understanding of user-item matching before the next turn $t+1$.

Through iterative reasoning and progressive evidence acquisition, the agent continuously reduces the information gap and refines its reasoning state. After $T$ steps, with sufficient user- and item-side evidence, RecThinker summarizes the inferred preferences and contextual knowledge to produce the final rank list.




\subsection{Tool Design and Information Acquisition}~\label{sec:Tools}
To support multi-step reasoning and bridge the information gaps in the reasoning process, we design a suite of specialized tools $\mathcal{A}$. These tools enable the agent to acquire complementary information from multiple perspectives and meet the information needs in reasoning. Specifically, we categorize the tools into three groups according to the information type: (1) user-side profiling tools, (2) item-side attribute tools, and (3) collaborative information tools.


\subsubsection{User-Side Information Acquisition.}
To construct a comprehensive image of the user’s preferences, we design two tools for user-side information retrieval: 
(1) \textbf{User Profile Search}: This tool retrieves static user attributes and long-term preferences, such as demographics, general interests, and behavior patterns.
It provides a summary of the user's preferences, serving as the foundation of preference reasoning.
(2) \textbf{User History Search}: It provides access to the user's interaction history. Each invocation returns some of the most recent items, together with detailed metadata and user feedback signals. 
To support progressive analysis, it can be called multiple times, enabling the agent to incrementally retrieve deeper historical context based on the complexity of the current task.

\subsubsection{Item-Side Information Acquisition.}
To enable fine-grained understanding of candidate or history items, we introduce the \textbf{Item Info Search} tool. It retrieves detailed attributes of a given item and expands its context using an Item Relation Graph, which identifies highly related items and ranks them based on a score function, reflecting co-occurrence patterns and categorical similarities.
By accessing detailed item attributes and structural relationships, the agent can analyze nuanced differences among similar candidates and match item characteristics with inferred user preferences. 

\subsubsection{Collaborative Information Acquisition.}
RecThinker further exploits collaborative information through two tools to support reasoning when a user's own data is limited:
(1) \textbf{Similar Users Search}: It retrieves users whose behavior patterns and profiles are most similar to the target user. It integrates sparse interaction co-occurrence statistics with dense profile embeddings to estimate user similarity. It is critical for fine-grained preference disambiguation and discovering potential interests not yet reflected in the target user's own history.
(2) \textbf{Knowledge Graph Search}: It leverages a knowledge graph to extract high-order collaborative evidence through multi-hop relational paths. By aggregating information from users and items connected through two-hop and three-hop paths, this tool provides high-order information and structural evidence to support ranking decisions in data-sparse scenarios.

\subsubsection{Progressive Tool Invocation Strategy.}
During reasoning, the agent dynamically selects and invokes tools according to its current reasoning state and information gap. Rather than exhaustively querying all sources, RecThinker adopts a progressive acquisition strategy: it first gathers coarse-grained user and item signals, and subsequently retrieves more fine-grained or collaborative evidence when ambiguity remains.
This flexible tool usage mechanism enables the agent to balance information sufficiency and efficiency, avoiding redundant queries while ensuring that critical evidence is available for accurate ranking.

\subsection{Two-Stage Training Strategy}
To ensure that RecThinker generates precise reasoning trajectories and invokes optimal tools, we propose a two-stage training paradigm that combines self-augmented SFT with RL. 
In the first stage, we generate and filter some high-quality trajectories and leverage them to conduct SFT, enabling the agent to autonomously reason over interaction chains and stabilize the agent's reasoning behavior and tool-invocation policy.
The second stage further refines the agent through GRPO on challenging instances, encouraging effective exploration of the reasoning space and improving its adaptability to complex recommendation scenarios.

\subsubsection{Trajectory Generation and Filtering}~\label{sec:data_construct}
Since raw trajectories may contain noisy reasoning or invalid tool usage, we adopt a generate-and-filter strategy to construct a set of reliable and high-quality reasoning trajectories to provide stable supervision for subsequent training.
Formally, given a training dataset $\mathcal{D}=\{(u, \mathcal{C}, y)\}$, where $y$ is the ranking label, we employ the base LLM to generate some reasoning trajectories. Each trajectory is represented as
\begin{equation}
\tau = (r_1, a_1, o_1, \dots, r_T, a_T),
\end{equation}
where the last action $a_T$ is the final ranking output.
To ensure the quality of supervision data, we evaluate each trajectory based on two criteria:
(1)~\textbf{ranking accuracy}, which is measured by whether the ground-truth item is ranked at the top position, 
and (2)~\textbf{format validity}, which verifies whether the trajectory strictly follows the predefined reasoning and tool-calling format.
Only trajectories that satisfy both criteria are retained:
\begin{equation}
\mathcal{T}_{\text{SFT}} = \{\tau \mid I_{\text{acc}}(\tau)=1 \land I_{\text{fmt}}(\tau)=1\},
\end{equation}
where $I_{\text{acc}}$ and $I_{\text{fmt}}$ denote the indicators of ranking correctness and format validity.
The high-confidence trajectory set $\mathcal{T}_{\text{SFT}}$ serves as the supervision data for the subsequent self-augmented SFT stage.

\subsubsection{Self-augmented Supervised Learning}~\label{sec:SFT}
Based on the filtered high-quality trajectories, we perform self-augmented SFT to stabilize the agent's policy and distill better reasoning patterns into the agent.
In this stage, each trajectory $\tau \in \mathcal{T}_{\text{SFT}}$ is transformed into a token sequence consisting of reasoning tokens, tool invocation tokens, tool observations, and final ranking outputs. During training, we supervise only the agent-generated tokens while masking environment responses.
Specifically, we partition each trajectory into two parts: 
\begin{equation}
\tau = (x^{\text{agent}}, x^{\text{env}}),
\end{equation}
where $x^{\text{agent}}$ includes reasoning, tool invocations, and final ranking outputs, and $x^{\text{env}}$ corresponds to observations.
We introduce a binary mask vector $\textbf{m} = (m_1,\dots,m_L)$ over the token sequence, where $m_i=1$ indicates agent-generated tokens and $m_i=0$ denotes tool-returned tokens, and $L$ is the sequence length. The learning objective is defined as: 
\begin{equation}
\mathcal{L}_{\text{SFT}}(\theta) = - \mathbb{E}_{\tau \sim \mathcal{T}_{\text{SFT}}} 
\sum_{i=1}^{L} m_i \log p_\theta(x_i \mid x_{<i}).
\end{equation}





\subsubsection{Policy Refinement via RL}~\label{sec:RL}
While SFT enables the agent to capture reliable reasoning patterns and acquire fundamental tool-use capabilities, it still suffers from limited exploration and suboptimal decision-making in complex scenarios.
To enhance exploration and robustness, we further optimize the agent using RL.

\textbf{Dataset Construction}. 
To construct effective RL samples, we adopt a difficulty-aware sampling strategy to focus on learnable yet challenging instances. For each sample not in the SFT training dataset, we perform multiple rollouts and evaluate their prediction correctness. Samples for which only a small portion of rollouts produce correct results are selected as hard cases, indicating that they are challenging but still solvable. These samples construct the RL training set $\mathcal{D}_{\mathrm{RL}}$.

\textbf{Reward Design}. We design a composite reward function to jointly encourage recommendation accuracy, format correctness, and tool utilization efficiency. 

(1) \textbf{Accuracy Reward}. We adopt NDCG@10 as the primary task reward to evaluate the accuracy of the generated ranking list. Given a trajectory $\tau$, the accuracy reward is defined as
\begin{equation}~\label{sec:AccReward}
R_{\mathrm{acc}}(\tau) = \mathrm{NDCG@10}(\tau).
\end{equation}

(2) \textbf{Format Reward}.
We introduce a binary format reward that evaluates whether the generated trajectory follows the predefined format, including the format of reasoning, tool calls, observations, and answer generation. Specifically, the format reward is defined as:
\begin{equation}~\label{sec:formatreward}
R_{\mathrm{fmt}}(\tau) =
\begin{cases}
0, & \text{if } \tau \text{ follows the predefined format}, \\
-1, & \text{otherwise}.
\end{cases}
\end{equation}

(3) \textbf{Tool Utilization Reward.}~\label{sec:toolreward}
To prevent insufficient reasoning (no tool calls) or redundant exploration (excessive calls), we define a piecewise linear reward function $R_{\mathrm{tool}}(\tau)$. If there is no tool call, we assign a constant penalty of $-1.0$. When the number of tool calls $1 \le N(\tau) \le 3$, the reward grows linearly as $R_{\mathrm{tool}}(\tau)= \frac{1}{3} N(\tau)$. For $3 \le N(\tau) \le 8$, the agent receives a maximum constant reward of 1.0 representing the best tool call range. When $N(\tau) > 8$, the reward begins to decay as $1.0 - \frac{3}{8}(N(\tau) - 8)$, and when $N(\tau) > 12$, it receives a great penlty of $R_{\mathrm{tool}}(\tau)=-0.5 - 0.3(N(\tau) - 12)$. 
The final reward is defined as a weighted combination:
\begin{equation}
R(\tau) = 
\begin{cases} 
-1, & \text{if } R_{\mathrm{fmt}}(\tau) = -1, \\
\lambda_1 R_{\mathrm{acc}}(\tau) + \lambda_2 R_{\mathrm{tool}}(\tau), & \text{otherwise},
\end{cases}
\end{equation}
where $\lambda_1, \lambda_2$ are hyperparameters.

\textbf{RL Algorithm}. We adopt GRPO~\cite{GRPO2024} as the RL algorithm due to its stability and efficiency in large-scale reasoning tasks. 
For each input $q$, we sample a group of $G$ trajectories $\{\tau_1, \dots, \tau_G\}$ from the current policy $\pi_\theta$. The GRPO objective is formulated as:
\begin{equation}
\begin{aligned}
\mathcal{L}_{\text{GRPO}}(\theta) &= - \mathbb{E}_{q \sim P(Q), \{o_i\}_{i=1}^G \sim \pi_{\theta_{old}}} \Bigg[ \frac{1}{G} \sum_{i=1}^G \\ & \bigg(  \min \left( \rho_i A_i, \text{clip}(\rho_i, 1-\epsilon, 1+\epsilon) A_i \right) 
 - \beta \mathbb{D}_{KL}(\pi_\theta \| \pi_{ref}) \bigg) \Bigg],
\end{aligned}\notag
\end{equation}
where $\rho_i = \frac{\pi_\theta(\tau_i \mid q)}{\pi_{\theta_{old}}(\tau_i \mid q)}$ is the importance sampling ratio. The advantage $A_i$ is computed by normalizing the rewards within the group.








\section{Experiments}
In this section, we evaluate the proposed RecThinker framework in different recommendation settings. Our experiments aim to address the following questions: 
\textbf{RQ1}: How does RecThinker perform compared to other traditional, LLM-based, and agent-based baselines? \textbf{RQ2}: How do different training stages affect the overall performance of RecThinker? \textbf{RQ3}: How do different reward designs contribute to RecThinker in the RL stage? \textbf{RQ4}: What are the individual contributions of different tool modules in RecThinker? \textbf{RQ5}: How does RecThinker generalize across different backbone model scales? \textbf{RQ6}: How do historical sequence length and data density impact the effectiveness of RecThinker?

\subsection{Experimental Setup}

\begin{table}[t]
\small
\centering
\caption{Statistics of the datasets.}
\setlength{\tabcolsep}{.6mm}{
\begin{tabular}{l@{}rrrrrr}
\toprule
\textbf{Datasets} & \textbf{\# Users} & \textbf{\# Items} & \textbf{\# Inters.} & \textbf{\# Inters/U.} & \textbf{\# Inters/I.} & \textbf{Sparsity} \\
\midrule
CDs & 93,653 & 64,031 & 1,178,439 & 12.58 & 18.40 & 99.98\% \\
\quad$\hookrightarrow$ Sparse & 100 & 704 & 800 & 8.00 & 1.14 & 98.86\% \\
\quad$\hookrightarrow$ Dense & 100 & 453 & 800 & 8.00 & 1.77 & 98.23\% \\
MovieLens & 6,040 & 3,883 & 1,000,209 & 165.60 & 257.59 & 95.74\% \\
\quad$\hookrightarrow$ Sparse & 100 & 1,880 & 5,000 & 50.00 & 2.66 & 97.34\% \\
\quad$\hookrightarrow$ Dense & 100 & 1,330 & 5,000 & 50.00 & 3.76 & 96.24\% \\
\bottomrule
\end{tabular}
}
\label{dataset}
\end{table}

\subsubsection{Dataset}
Following~\cite{AgentCF2023, PersonaX2025}, we conduct experiments on two widely used datasets: Amazon CD \& Vinyl and MovieLens-1M. For each dataset, following~\cite{AgentCF2023}, we construct two subsets with different data densities for evaluation, each containing 100 users. Statistics of the datasets are shown in Table~\ref{dataset}. More details are in Appendix~\ref{dataset details}.

\begin{table*}[t]
\small
\centering
\caption{The recommendation results of NDCG@1, 5, 10 on four datasets. The best results are shown in bold.}
\setlength{\tabcolsep}{6.pt}{
\begin{tabular}{lcccccccccccc}
\toprule
{\multirow{2}{*}{Method}} & \multicolumn{3}{c}{$\text{CDs}_{\text{sparse}}$} & \multicolumn{3}{c}{$\text{CDs}_{\text{dense}}$} & \multicolumn{3}{c}{$\text{MovieLens}_{\text{sparse}}$} & \multicolumn{3}{c}{$\text{MovieLens}_{\text{dense}}$} \\
\cmidrule(lr){2-4}\cmidrule(lr){5-7}\cmidrule(lr){8-10}\cmidrule(lr){11-13}    
\multicolumn{1}{c}{} & N@1 & N@5 & N@10 & N@1 & N@5 & N@10 & N@1 & N@5 & N@10 & N@1 & N@5 & N@10 \\
\midrule
$\text{BPR}_{\text{sample}}$ & 0.1200 & 0.3291 & 0.4514 & 0.2100 & 0.3063 & 0.4904 & 0.1700 & 0.3786 & 0.4556 & 0.2400 & 0.4226 & 0.4854 \\
$\text{SASRec}_{\text{sample}}$ & 0.1900 & 0.3910 & 0.5050 & 0.2600 & 0.3959 & 0.5254 & 0.2300 & 0.4301 & 0.5279 & 0.3200 & 0.4904 & 0.5526 \\
Pop & 0.0900 & 0.2502 & 0.3988 & 0.1100 & 0.2608 & 0.4216 & 0.1300 & 0.2913 & 0.4059 & 0.1500 & 0.3530 & 0.4245 \\
LLMSeqSim & 0.1200 & 0.3014 & 0.4512 & 0.2100 & 0.3899 & 0.5105 & 0.1600 & 0.3354 & 0.4485 & 0.2100 & 0.4029 & 0.4685 \\
LLMRank & 0.1400 & 0.3134 & 0.4656 & 0.2700 & 0.4082 & 0.5263 & 0.2000 & 0.4010 & 0.4915 & 0.2600 & 0.4683 & 0.5171 \\
R2Rec & \underline{0.2700} & \underline{0.4613} & \underline{0.5468} & \underline{0.3500} & \underline{0.5314} & \underline{0.5970} & \underline{0.2600} & \underline{0.4760} & 0.5303 & \underline{0.3500} & \underline{0.5200} & 0.5910 \\
AgentCF & 0.2400 & 0.4328 & 0.5527 & \underline{0.3500} & 0.4737 & 0.5739 & 0.2400 & 0.4479 & 0.5437 & 0.3400 & 0.5007 & 0.5882 \\
PersonaX & 0.2400 & 0.4439 & 0.5332 & 0.3300 & 0.5119 & 0.5868 & 0.2500 & 0.4637 & \underline{0.5542} & 0.3300 & 0.5119 & \underline{0.5921} \\
\rowcolor[RGB]{236,244,252}RecThinker & \textbf{0.3100} & \textbf{0.5365} & \textbf{0.6174} & \textbf{0.3900} & \textbf{0.5711} & \textbf{0.6601} & \textbf{0.2900} & \textbf{0.5228} & \textbf{0.5964} & \textbf{0.4000} & \textbf{0.5821} & \textbf{0.6619} \\
\rowcolor[RGB]{236,244,252}\textit{Improvement} & +14.81\% & +16.30\% & +11.71\% & +11.43\% & +7.47\% & +10.57\% & +11.54\% & +9.83\% & +7.61\% & +14.29\% & +11.94\% & +11.79\% \\
\bottomrule
\end{tabular}
}
\label{main_result}
\end{table*}

\subsubsection{Baseline Models}
We select eight representative methods as baselines, including one ad-hoc model Pop, two traditional recommendation models BPR-MF~\cite{BPR2012} and SASRec~\cite{SASRec2018}, two LLM-based methods LLMSeqSim~\cite{LLMSeqSim2023} and LLMRank~\cite{LLMRank2024}, one reasoning method R2Rec~\cite{Reason2Recommend2025}, and two agentic recommendation methods AgentCF~\cite{AgentCF2023} and PersonaX~\cite{PersonaX2025}.  
Detailed descriptions of these models are in Appendix~\ref{append_baseline}.

\begin{table*}[t]
\small
\centering
\caption{Performance (NDCG@1, 5, 10) of ablation models on four datasets.}
\setlength{\tabcolsep}{6.6pt}{
\begin{tabular}{lcccccccccccc}
\toprule
{\multirow{2}{*}{Method}} & \multicolumn{3}{c}{$\text{CDs}_{\text{sparse}}$} & \multicolumn{3}{c}{$\text{CDs}_{\text{dense}}$} & \multicolumn{3}{c}{$\text{MovieLens}_{\text{sparse}}$} & \multicolumn{3}{c}{$\text{MovieLens}_{\text{dense}}$} \\
\cmidrule(lr){2-4}\cmidrule(lr){5-7}\cmidrule(lr){8-10}\cmidrule(lr){11-13}    
 & N@1 & N@5 & N@10 & N@1 & N@5 & N@10 & N@1 & N@5 & N@10 & N@1 & N@5 & N@10 \\
 \midrule
\rowcolor[RGB]{236,244,252}RecThinker & \textbf{0.3100} & \textbf{0.5365} & \textbf{0.6174} & \textbf{0.3900} & \textbf{0.5711} & \textbf{0.6601} & \textbf{0.2900} & \textbf{0.5228} & \textbf{0.5964} & \textbf{0.4000} & \textbf{0.5821} & \textbf{0.6619} \\
\midrule
\quad \textit{w/o.} SFT & 0.2800 & 0.4598 & 0.5486 & 0.3500 & 0.5423 & 0.6278 & 0.2700 & 0.4956 & 0.5780 & 0.3500 & 0.5420 & 0.6154 \\
\quad \textit{w/o.} RL & 0.2700 & 0.5159 & 0.5954 & 0.3700 & 0.5493 & 0.6329 & 0.2800 & 0.5007 & 0.5820 & 0.3600 & 0.5570 & 0.6386 \\
\quad \textit{w/o.} SFT \& RL & 0.2500 & 0.4617 & 0.5510 & 0.3200 & 0.5179 & 0.6089 & 0.2500 & 0.4796 & 0.5582 & 0.3400 & 0.5377 & 0.5950 \\
\midrule
\quad \textit{w/o.} Acc Reward & 0.2700 & 0.4829 & 0.5705 & 0.3600 & 0.5461 & 0.6290 & 0.2700 & 0.5051 & 0.5715 & 0.3600 & 0.5441 & 0.6226 \\
\quad \textit{w/o.} Fmt Reward & 0.2500 & 0.4746 & 0.5653 & 0.3700 & 0.5495 & 0.6321 & 0.2600 & 0.4917 & 0.5687 & 0.3600 & 0.5518 & 0.6293 \\
\quad \textit{w/o.} Tool Reward & 0.2800 & 0.5034 & 0.5883 & 0.3800 & 0.5675 & 0.6435 & 0.2800 & 0.5143 & 0.5901 & 0.3800 & 0.5637 & 0.6380 \\
\midrule
\quad \textit{w/o.} Profile Tool & 0.2700 & 0.4978 & 0.5773 & 0.3700 & 0.5541 & 0.6356 & 0.2800 & 0.5014 & 0.5727 & 0.3600 & 0.5478 & 0.6165 \\
\quad \textit{w/o.} Hist Tool & 0.2200 & 0.4271 & 0.5096 & 0.3100 & 0.5122 & 0.5945 & 0.2400 & 0.4116 & 0.5261 & 0.3000 & 0.5096 & 0.5810 \\
\quad \textit{w/o.} SimU Tool & 0.2800 & 0.5108 & 0.5796 & 0.3700 & 0.5472 & 0.6301 & 0.2700 & 0.4736 & 0.5698 & 0.3600 & 0.5591 & 0.6346 \\
\quad \textit{w/o.} Item Tool & 0.2400 & 0.4720 & 0.5549 & 0.3500 & 0.5413 & 0.6276 & 0.2500 & 0.4743 & 0.5624 & 0.3500 & 0.5312 & 0.6022 \\
\quad \textit{w/o.} KG Tool & 0.2900 & 0.5141 & 0.5912 & 0.3800 & 0.5642 & 0.6343 & 0.2800 & 0.5078 & 0.5832 & 0.3700 & 0.5583 & 0.6403 \\
\bottomrule
\end{tabular}
}
\label{ablation}
\end{table*}

\subsubsection{Evaluation Metrics}
We evaluate the performance of all methods using NDCG@$k$ ($k\in\{1,5,10\}$), which is widely used in recommendation~\cite{AgentCF2023, PersonaX2025}. 
Following previous works~\cite{AgentCF2023}, we adopt the leave-one-out strategy for evaluation. 
For each test sample, we randomly select nine negative items and combine them with the label item to construct the candidate set. To ensure fair comparisons, we pre-sample and store the candidate sets in advance and evaluate all methods on the same ranking lists.

\subsubsection{Implementation Details}
In our experiments, we adopt QWQ-32B~\cite{QWen25_QWQ} as the backbone LLM for our agent. The temperature is set to 1.0, and the top-p value is set to 0.95. 
The prompts contain detailed instructions, descriptions of each tool, user id, and candidate items.
During the SFT stage, we apply Low-Rank Adaptation (LoRA)~\cite{LoRA2022} to fine-tune the backbone model, with a learning rate of $1e^{-5}$. We use 8,000 training samples and train the model for 3 epochs.
During the RL stage, we optimize the SFT model using LoRA, with a learning rate of $1e^{-6}$. We train the model on 1,000 hard samples for 5 epochs. We set the number of trajectories $G$ in GRPO as 8. The weights of the reward function $\lambda_1, \lambda_2$ are set to 1.0 and 0.1.
At the evaluation stage, each test sample is evaluated multiple times, and the average results are reported.
For reproducibility, we release our code at \url{https://github.com/Aska-zhang/RecThinker}.




\subsection{Main Results (RQ1)}
The recommendation performance of RecThinker and all baseline methods is reported in Table~\ref{main_result}. Our key findings are as follows: 

(1) RecThinker consistently outperforms all baseline methods on all datasets. In particular, it achieves significant performance improvements of 11.71\%, 10.57\%, 7.61\%, and 11.79\% over the strongest baseline on the four datasets and the NDCG@10 metric, respectively. It demonstrates the effectiveness of RecThinker in autonomously analyzing information gaps, formulating reasoning plans, and acquiring diverse knowledge to enhance the recommendation process.

(2) RecThinker achieves better performance than traditional recommendation models. It is because RecThinker can integrate collaborative signals with external user and item knowledge and autonomously analyze and incorporate additional information into the ranking and recommendation process, leading to more precise user-item matching.

(3) Compared with existing agent-based and LLM-based methods, RecThinker achieves superior performance across all evaluation settings. It indicates that our method can effectively analyze the information gap, obtain more informative evidence, and conduct accurate reasoning, which enables precise information acquisition and reliable recommendations. 
It also demonstrates the efficacy of the two-stage optimization in refining the model's policy for more efficient exploration.

(4) Agent-based methods and reasoning methods generally outperform traditional and other LLM-based models under the same training data settings. This means that explicitly modeling the recommendation process through agentic frameworks and reasoning can capture nuanced user preferences and item characteristics and facilitate better user-item matching and ranking performance.

\subsection{Effects of Training Stage (RQ2)}
To investigate the effectiveness of the two training stages in RecThinker, we conduct a series of ablation studies:
(1) \textbf{w/o. SFT}: We remove the self-enhanced SFT stage in Section~\ref{sec:SFT}, and directly perform RL for training and evaluation.
(2) \textbf{w/o. RL}: We remove the RL stage described in Section~\ref{sec:RL}, and only perform SFT for training and evaluation.
(3) \textbf{w/o. SFT \& RL}: We remove the SFT and RL stage, and directly perform evaluation with the base LLM.

The results of the ablation studies are reported in Table~\ref{ablation}. We find: 
(1) The performance degradation of \textbf{w/o. SFT} and \textbf{w/o. SFT \& RL} indicates that the self-augmented SFT stage can effectively improve the reasoning abilities of agents and stabilize the learning policy. Moreover, it facilitates the learning of high-quality reasoning patterns and output formats, thereby providing an effective warm-up for subsequent RL.
(2) RecThinker consistently outperforms \textbf{w/o. RL} and \textbf{w/o. SFT \& RL}, demonstrating that RL further enhances the exploration capability of the model under format constraints. In particular, it enables the model to perform more effective reasoning and analysis in complex scenarios.

\subsection{Reward Score Analysis (RQ3)}
To quantify the contribution of the three reward scores in the RL stage, we conduct some ablation studies by removing each score: 
(1) \textbf{w/o. Acc Reward}: We remove the accuracy reward function in Equation~(\ref{sec:AccReward}) during the RL stage.
(2) \textbf{w/o. Fmt Reward}: We remove the format reward in Equation~(\ref{sec:formatreward}) during the RL stage.
(3) \textbf{w/o. Tool Reward}: We remove the tool utilization reward in Section~\ref{sec:toolreward} during the RL stage.
The results of the ablation studies are reported in Table~\ref{ablation}. We can observe the following findings: 
The low result of \textbf{w/o. Acc Reward} suggests that the accuracy reward serves as the most critical guidance signal, which provides clear optimization directions for recommendations. The performance decline of \textbf{w/o. Fmt Reward} indicates that format constraints play an essential role in preventing ineffective exploration and invalid generation. In addition, \textbf{w/o. Tool Reward} demonstrates that proper tool usage is important for acquiring sufficient yet non-redundant information.


\subsection{Tool Analysis (RQ4)}
In this section, we analyze the utilization of different tools introduced in Section~\ref{sec:Tools}, as well as their impact on the final recommendation performance. Specifically, we investigate the invocation frequency of each tool during inference and conduct some ablation experiments \textbf{w/o. Profile Tool}, \textbf{w/o. Hist Tool}, \textbf{w/o. SimU Tool}, \textbf{w/o. Item Tool}, and \textbf{w/o. KG Tool} by removing each tool respectively.
The usage frequency of each tool is illustrated in Table~\ref{Tool_analysis}, and the corresponding ablation results are reported in Table~\ref{ablation}.
As shown in Table~\ref{Tool_analysis}, the Profile Tool is the most frequently used tool, indicating that user profiles are the foundation of the ranking and reasoning process. 
The History Tool and Item Tool are the second most frequently invoked tools. These tools provide detailed historical behaviors and fine-grained item attributes, which offer rich contextual information for user behavior modeling and precise user-item matching.
The Similar User Search Tool and KG Tool are invoked occasionally but still provide performance improvements. 
They complement the reasoning process by introducing auxiliary information of collaborative and structural knowledge, which further refines the matching results in complex scenarios.
The ablation results in Table~\ref{ablation} further confirm the importance of these tools. Removing any individual tool leads to consistent performance degradation, demonstrating that each tool contributes to the overall reasoning capability of RecThinker. In particular, the removal of the History Tool and Item Tool causes the most significant performance drop, highlighting their critical role in establishing accurate user image and item semantics.


\begin{figure}[t]
\centering
\includegraphics[width=1.0\linewidth]{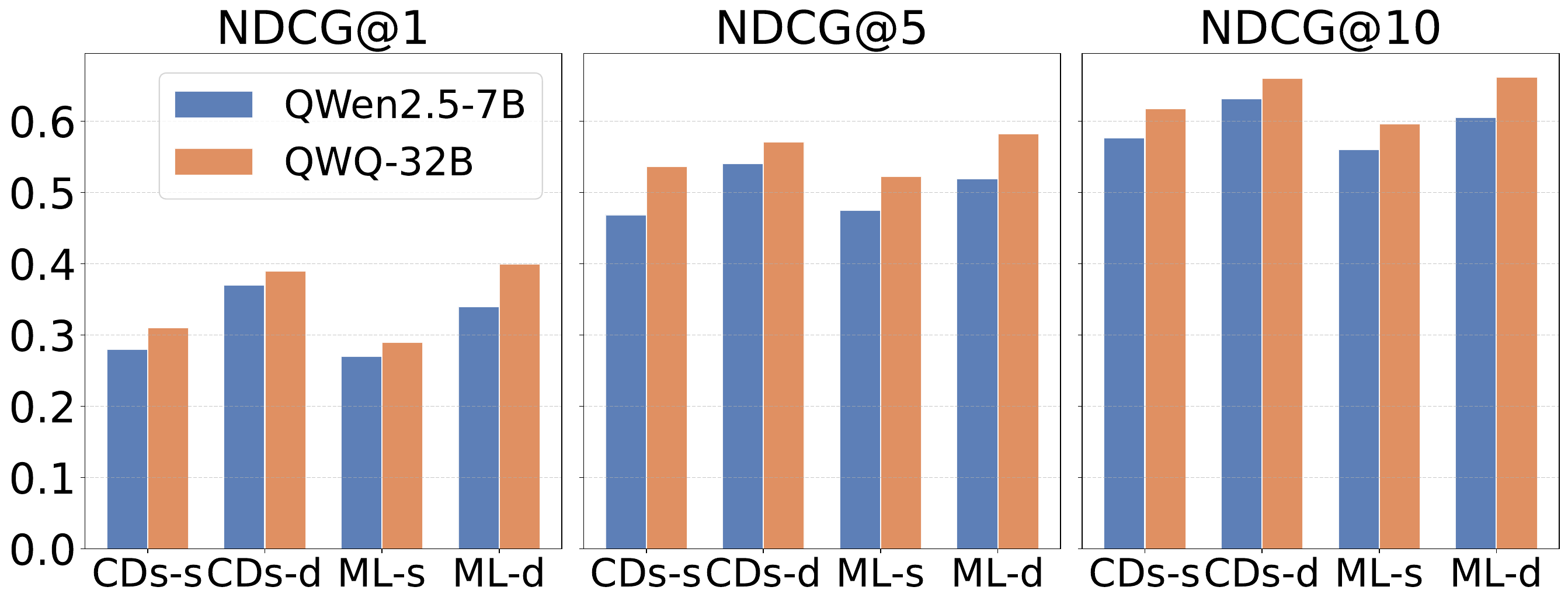}
\caption{NDCG@1, 5, 10 results of RecThinker with QWQ-32B and QWen2.5-7B as backbone model on four datasets. CDs-s, CDs-d, ML-s, ML-d represents $\text{CDs}_{\text{sparse}}$, $\text{CDs}_{\text{dense}}$, $\text{MovieLens}_{\text{sparse}}$, $\text{MovieLens}_{\text{dense}}$, respectively. }
\label{backbone}
\end{figure}

\begin{table}[t]
\small
\centering
\caption{Five tool usage frequency on four datasets.}
\setlength{\tabcolsep}{4pt}{
\begin{tabular}{lcccc}
\toprule
Tool & $\text{CDs}_{\text{sparse}}$ & $\text{CDs}_{\text{dense}}$ & $\text{MovieLens}_{\text{sparse}}$ & $\text{MovieLens}_{\text{dense}}$ \\
\midrule
Profile Tool & 97.00\% & 95.00\% & 98.00\% & 94.00\% \\
History Tool & 89.00\% & 91.00\% & 94.00\% & 93.00\% \\
SimU Tool & 24.00\% & 17.00\% & 20.00\% & 19.00\% \\
Item Tool & 70.00\% & 61.00\% & 90.00\% & 79.00\% \\
KG Tool & 15.00\% & 12.00\% & 13.00\% & 11.00\% \\
\bottomrule
\end{tabular}
}
\label{Tool_analysis}
\end{table}

\subsection{Backbone Model Analysis (RQ5)}
In this section, we investigate the impact of different backbone models on the performance of RecThinker. We replace the backbone models used in the SFT and RL stages with Qwen2.5-7B, and still use QWQ-32B for data construction in Section~\ref{sec:data_construct}. 
This setting allows us to evaluate whether the proposed reasoning and training paradigm can be effectively transferred to other backbone models.
The experimental results in Figure~\ref{backbone} show that RecThinker equipped with Qwen2.5-7B still achieves competitive performance across all datasets. Although a moderate performance gap exists compared with the QWQ-32B backbone, the trained model outperforms LLM-based and agentic baselines on most of the metrics, demonstrating that our proposed paradigm is not restricted to a specific backbone.
These results indicate that RecThinker exhibits good generalizability and scalability. By leveraging high-quality reasoning supervision and RL, smaller backbones can also learn effective reasoning patterns and tool-usage strategies, leading to substantial performance improvements in recommendation tasks.

\begin{figure}[t]
\centering
\includegraphics[width=1.0\linewidth]{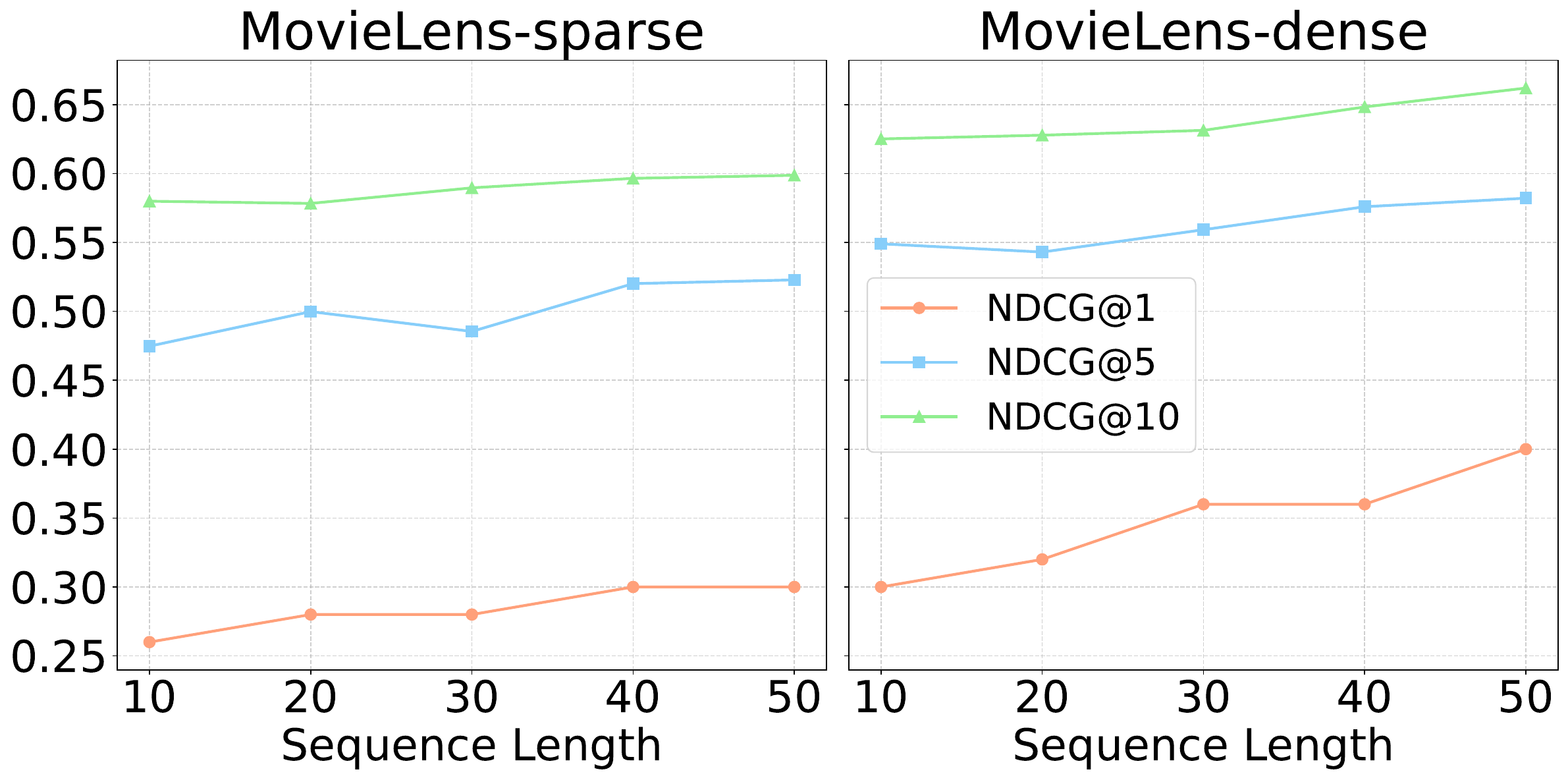}
\caption{NDCG@1, 5, 10 results of RecThinker with different sequence length on two MovieLens-1M datasets.  }
\label{sequence}
\end{figure}

\subsection{Impact of Sequence Length (RQ6)}
In this section, we investigate how user sequence length and interaction density affect the performance of RecThinker. 
Specifically, we construct ten subsets from the MovieLens-1M dataset with the same users and consider two factors: interaction density (dense and sparse) and sequence length (from 10 to 50). 
We then evaluate the performance of the trained model on these subsets.
The experimental results are shown in Figure~\ref{sequence}. We observe that RecThinker achieves better performance on datasets with longer user sequences, indicating that richer historical information enables more accurate preference modeling and reasoning. Moreover, the performance gain from longer sequences is more pronounced in dense settings, indicating that increasing data size is important for dense datasets.
These results demonstrate that RecThinker can effectively leverage extended user histories to obtain more information and achieve better reasoning. 

\section{Conclusion}
In this work, we propose {RecThinker}, an agentic framework for tool-augmented reasoning in recommendation. It dynamically plans the reasoning paths, analyzes information requirements, and invokes external tools actively based on the information needs.
Specifically, we design an Analyze-Plan-Act reasoning paradigm that enables the model to assess the sufficiency of user- and item-side evidence, autonomously invoke tool calls, and progressively refine its reasoning process. To support effective information acquisition, we construct a set of specialized tools for user information, item details, and collaborative signals acquisition in recommendation scenarios.
Moreover, we introduce a two-stage self-augmented training strategy that leverages high-quality trajectories generated by the backbone to train the agent through SFT and RL and improve the reasoning accuracy and tool invocation effectiveness. 
Extensive experiments on multiple benchmarks demonstrate the effectiveness and generalization of RecThinker across diverse recommendation scenarios.



\bibliographystyle{ACM-Reference-Format}
\bibliography{sample-base}

\appendix

\section{Details of Datasets}
\label{dataset details}
Following previous works~\cite{AgentCF2023, PersonaX2025}, we conduct experiments on two widely used datasets: Amazon CD \& Vinyl and MovieLens-1M. 
(1) Amazon is a large-scale product purchasing dataset, which consists of user-item interactions, item ratings, and item metadata.\footnote{Amazon dataset, \url{https://nijianmo.github.io/amazon/index.html}}
(2) MovieLens-1M is a widely adopted movie recommendation benchmark, containing user-movie ratings and corresponding movie metadata.\footnote{MovieLens-1M, \url{https://grouplens.org/datasets/movielens/1m/}}
For each dataset, following AgentCF~\cite{AgentCF2023} , we construct two subsets with different data densities for evaluation, considering different levels of data sparsity and sequential interaction lengths. Specifically, for each dataset, we construct two subsets with different interaction densities, and each subset contains 100 users. 
We treat interactions with ratings greater than 3 as positive feedback and retain them for training and evaluation. We adopt the leave-one-out strategy to construct the training, validation, and test sets.
The statistics of the datasets are summarized in Table~\ref{dataset}.

\section{Detailed description of baselines}
\label{append_baseline}
We select eight representative methods as baselines, covering ad-hoc models, traditional recommendation models, LLM-based methods, and agentic recommendation methods:

$\bullet$ \textbf{Pop}: It ranks items based on their popularity, measured by the total number of interactions.

$\bullet$ \textbf{BPR-MF}~\cite{BPR2012}: It is a classic collaborative filtering method that decomposes the user-item interaction matrix into latent representations and optimizes them with the BPR loss.

$\bullet$ \textbf{SASRec}~\cite{SASRec2018}: It is a representative sequential recommender that adopts a Transformer architecture with self-attention and feed-forward networks to model users' historical interaction sequences.

$\bullet$ \textbf{LLMSeqSim}~\cite{LLMSeqSim2023}: It is an LLM-based framework that encodes item textual metadata into embeddings using LLMs, and ranks candidate items by computing the similarity between session representations and target item embeddings.

$\bullet$ \textbf{LLMRank}~\cite{LLMRank2024}: It is a prompt-based ranking method that incorporates users' historical interactions into carefully designed prompts, and leverages in-context learning to employ large language models as rankers for candidate ranking.

$\bullet$ \textbf{R2Rec}~\cite{Reason2Recommend2025}: It is a reasoning-enhanced framework that constructs interaction chains and integrates them into the instructions to perform interaction-of-thought for LLM. It also designs an SFT and an RL stage to train the LLM with GPT-4omini as a teacher.

$\bullet$ \textbf{AgentCF}~\cite{AgentCF2023}: It is an agentic recommendation framework that models users and items as agents. It simulates interactions among agents based on training data to learn collaborative filtering knowledge, and performs ranking by measuring similarities between the learned agent memories.

$\bullet$ \textbf{PersonaX}~\cite{PersonaX2025}: It is an agent-based framework focused on dynamic user profiling. It clusters user historical interactions, dynamically selects representative sub-behaviors, and aggregates them into personalized profiles using LLMs for similarity-based ranking.

\section{Detailed Description of Toolsets}
\label{Toolset details}
In this section, we provide a detailed description of the tools we introduced in Section~\ref{sec:Tools}, including their construction, mechanisms, and returned contents.

\textbf{User Profile Search}: It retrieves a compact semantic profile summarizing the user's long-term preferences.
Specifically, for each user $u$, we take the user's historical interactions $S_u$ from the training set as input, and construct a profile generation prompt $\mathcal{P}_{\text{profile}}$ and feed it into an LLM to summarize the user's long-term preferences and behavioral patterns.:
\begin{equation}
p_u^{\text{sum}} = \text{LLM}(\mathcal{P}_{\text{profile}}, S_u),
\end{equation}
where $p_u^{\text{sum}}$ is a concise summary of user preferences.
For datasets containing explicit demographic information (e.g., MovieLens), we concatenate the demographic attributes $D_u$ (age, gender, occupation) with $p_u^{\text{sum}}$. The final profile is represented as $p_u = [D_u; p_u^{\text{sum}}]$, which is the output of this tool.


\textbf{User History Search}: It retrieves the chronological interaction history of a user. For each user $u$, the history is organized in chronological order and represented as $S_u = \{i_{1},..., i_{n}\}$. Upon invocation, the tool retrieves the $k$ most recent items $\{i_{n-k+1},..., i_{n}\}$ and returns their basic information (e.g., title and category). 
To support progressive exploration of long-term preferences, the tool allows iterative backward retrieval. If the agent invokes the tool multiple times, each subsequent call returns the previous $k$ interactions and their basic information before the last retrieved segment:
\begin{equation}
\mathcal{S}_u^{(m)} = \{ i_{n-mk+1}, \dots, i_{n-(m-1)k} \},
\end{equation}
where $m$ is the invocation step. Meanwhile, the tool also outputs a binary indicator $\delta_u$ specifying whether the complete history has been retrieved.
This design enables the agent to flexibly control the granularity of historical information based on task demands.


\textbf{Item Info Search}: It retrieves detailed metadata and relational information for a given item, which can be used for both historical and candidate items.

For each item, the tool first retrieves detailed attributes $A_i$ such as title, category, brand, price, and textual descriptions for Amazon datasets, or genre and release year for movie datasets. We can also use LLM to expand its knowledge and use it in the retrieval (not used in this paper). 

In addition, to complement the information of the item, we further construct an Item-Relation Graph $G_I=(\mathcal{E},\mathcal{R})$ based on multiple relations to capture the information of related items. For instance, in the Amazon dataset, the entities $\mathcal{E}$ in this graph include items, categories, and brands. Relations $\mathcal{R}$ include \textit{also\_bought}, \textit{also\_viewed}, \textit{belong\_to\_category}, and \textit{produced\_by\_brand}. For a target item $i$, it finds some related items through path-based retrieval on this graph, in which the relations between two items can be categorized into four types: \textit{also\_bought}, \textit{also\_viewed}, \textit{same\_category}, and \textit{same\_brand}. Because of the large space of the retrieval space, we define a scoring mechanism and assign weights $w_r$ to each relation $r$ reflecting its semantic strength:
\begin{equation}
w_r =
\begin{cases}
3, & r=\text{also-bought},\\
2, & r=\text{also-viewed},\\
1, & r \in \{\text{same-category}, \text{same-brand}\}.
\end{cases}
\end{equation}
We think that the relation \textit{also\_bought} is a strong signal for related items, which reveals that users will buy these two items together or be interested in the related item. For a target item $i$, the similarity score with a related item $j$ is calculated as the sum of weights of all relations $\mathcal{R}_{i,j}$ connecting them:
\begin{equation}
s(i,j) = \sum_{r \in \mathcal{R}_{ij}} w_r,
\end{equation}
where $\mathcal{R}_{ij}$ denotes all relations between $i$ and $j$. We rank related items by $s(i,j)$ and return the top-$K$ neighbors:
\begin{equation}
\mathcal{N}_i = \text{TopK}_j \; s(i,j).
\end{equation}
Finally, this tool returns the detailed attributes $A_i$ of the target item $i$, and the basic information of the top-$K$ items and the relations between the target item and the related items as supplementary context.


\textbf{Similar Users Search}: It identifies users with similar preferences to the target user and retrieves their profiles for preference enrichment.
We adopt a hybrid similarity computation strategy combining sparse and dense signals for similar user retrieval.

(1) \textbf{Sparse similarity}. We construct a user-item co-occurrence matrix and compute cosine-style similarity:
\begin{equation}
s_{\text{sparse}}(u,v)=\frac{|I_u \cap I_v|}{\sqrt{|I_u||I_v|}},
\end{equation}
where $I_u$ and $I_v$ denote the interacted item sets of users $u$ and $v$, respectively.

(2) \textbf{Dense similarity}. We use a pre-trained encoder model BGE-m3~\cite{BGEm3} to encode user profiles and compute the cosine similarity of the embeddings between two users: 
\begin{equation}
\begin{aligned}
    \mathbf{e}_u &= \text{Enc}(\text{Profile}(u)), \\
    s_{\text{dense}}(u, v) &= \text{cos}(\mathbf{e}_u, \mathbf{e}_v).
\end{aligned}
\end{equation}

(3) \textbf{Hybrid Similarity}: The final similarity score is obtained by weighted aggregation:
\begin{equation}
s_{\text{hybrid}}(u,v)=\alpha s_{\text{sparse}}(u,v)+(1-\alpha)s_{\text{dense}}(u,v),
\end{equation}
The tool returns the top-$k$ most similar users ranked by $s_{\text{hybrid}}(u,v)$ and their corresponding profiles as collaborative information for better reasoning.


\textbf{Knowledge Graph Search}: It retrieves high-order relational information from a User-Item knowledge graph to supplement collaborative information.

We construct a graph $\mathcal{G}_K=(\mathcal{E}_K,\mathcal{R}_K)$, 
where $\mathcal{E}_K$ includes users, items, and attributes, and $\mathcal{E}_K$ represents relations between entities. 
Starting from target user $u$, the tool performs two-hop and three-hop traversals to identify related users and items (for example, $u \xrightarrow{buy} i \xrightarrow{bought\_by} v$; $u \xrightarrow{buy} i \xrightarrow{also\_bought} i' \xrightarrow{buy} v$). The retrieved paths are converted into interpretable explanations. For example, the tool may return the profile of a related user $v$ together with an explanation such as "User $u$ may have the same interests as $v$ because both $u$ and $v$ purchased item $i$ (or share the same age, gender, and occupation group)." or "User $u$ may have the same interests as $v$ because $u$ purchased item $i$, $v$ purchased item $i'$, $i$ and $i'$ are often bought together."

It randomly selects $k_1$ two-hop paths and $k_2$ three-hop paths ($k_1 < k_2$), and returns the profiles of the end user node of the paths together with the explanation sentences. By introducing multi-hop relational structures, this tool provides more structural and collaborative knowledge of users and items, enabling better reasoning and recommendations.



\end{document}